\documentclass[a4paper, 10pt, conference]{ieeeconf}      

\IEEEoverridecommandlockouts                              

\usepackage{fullpage}
\usepackage{graphicx}
\usepackage{amsmath}
\usepackage{amssymb}
\usepackage{amsthm}
\usepackage{authblk}
\usepackage{hyperref}
\usepackage{verbatim}
\usepackage{algorithm2e}

\newcommand{\R}{\mathbb{R}}

\newcommand{\N}{\mathbb{N}}

\newcommand{\Nmax}{N_{\max}}
\newcommand{\Nminers}{N_{\text{net}}}
\newcommand{\Nshards}{N_{\text{shards}}}

\newcommand{\compat}{\mathsf{compat}}

\usepackage[english]{babel}
\usepackage[utf8]{inputenc}

\newcommand{\Gminer}{G_{\text{network}}}
\newcommand{\Vminer}{V_{\text{network}}}
\newcommand{\Eminer}{E_{\text{network}}}
\newcommand{\Wminer}{W_{\text{network}}}

\newcommand{\Gbase}{G_{\text{base}}}
\newcommand{\Vbase}{V_{\text{base}}}
\newcommand{\Ebase}{E_{\text{base}}}

\newcommand{\Chainpred}{\Lambda_{\text{Chainweb}}}
\newcommand{\GenesisTree}{\mathcal{T}_{\emptyset}}

\newcommand{\Expect}{\mathsf{E}}
\newcommand{\Var}{\mathsf{Var}}
\newcommand{\height}{\mathsf{height}}

\newtheorem{defn}{Definition}
\newtheorem{claim}{Claim}

\title{\LARGE \bf
Agent-Based Simulations of Blockchain protocols illustrated via Kadena's \emph{Chainweb}
}

\author{Tarun Chitra$^{1}$, \thanks{$^{1}$T. Chitra is at Gauntlet Networks, Inc. \\
        {\tt\small tarun@gauntlet.network}}%
        Monica Quaintance$^{2}$, Stuart Haber$^{3}$, and Will Martino$^{4}$ \\
\thanks{$^{2}$M. Quaintance is at Kadena LLC \\
        {\tt\small monica@kadena.io}}%
\thanks{$^{3}$S. Haber is Stuart Haber Crypto LLC \\ 
        {\tt\small stuart.haber@acm.org}}%
\thanks{$^{3}$W. Martino is at Kadena LLC \\
        {\tt\small will@kadena.io}}%
}

\begin{document}

\maketitle
\begin{abstract}
   While many distributed consensus protocols provide robust liveness and consistency guarantees under the presence of malicious actors, quantitative estimates of how economic incentives affect security are few and far between. In this paper, we describe a system for simulating how adversarial agents, both economically rational and Byzantine, interact with a blockchain protocol. This system provides statistical estimates for the economic difficulty of an attack and how the presence of certain actors influences protocol-level statistics, such as the expected time to regain liveness. This simulation system is influenced by the design of algorithmic trading and reinforcement learning systems that use explicit modeling of an agent's reward mechanism to evaluate and optimize a fully autonomous agent. We implement and apply this simulation framework to Kadena's Chainweb, a parallelized Proof-of-Work system, that contains complexity in how miner incentive compliance affects security and censorship resistance. We provide the first formal description of Chainweb that is in the literature and use this formal description to motivate our simulation design. Our simulation results include a phase transition in block height growth rate as a function of shard connectivity and empirical evidence that censorship in Chainweb is too costly for rational miners to engage in. We conclude with an outlook on how simulation can guide and optimize protocol development in a variety of contexts, including Proof-of-Stake parameter optimization and peer-to-peer networking design.
\end{abstract}

\section{Introduction}
Blockchain systems provide security via the economic disincentivization of malicious behaviors, such as double-spending or long-range attacks. Formally, this security is established by proving that the system achieves liveness, consistency, and persistence under suitable networking conditions (e.g. partial synchrony or asynchrony) and under the assumption that honest and rational agents are in the majority \cite{pass2017analysis, gilad2017algorand, hanke2018dfinity}. Many of these results give precise quantitative conditions that prescribe when a protocol will fail as a function of parameters, such as the difficulty adjustment period \cite{pass2017analysis} or epoch length \cite{gilad2017algorand}. However, it becomes exceedingly difficult to provide similarly strong results when dealing with sharded or parallelized blockchains. There is a marked loss in quantitative strength in the existing literature of results, as accommodating for communication complexity and resource (e.g. energy, space, or stake) redistribution leads to significantly worse liveness performance \cite{zamani2018rapidchain, kokoris2018omniledger} and weaker tolerances for Byzantine actors. Mathematically, the main reason for this loss in statistical power comes from an inability to apply non-trivial concentration inequalities to all chains simultaneously, due to residual correlations that stem from cross-shard transactions and the overhead of repeated committee selection. Moreover, the increased communication cost for cross-shard or cross-chain transactions is only described asymptotically, which makes it hard to tune practical peer-to-peer networking algorithms for cross-shard transactions. Finally, as one increases the number of shards, there is often a dramatic increase in the number of parameters, such as timeouts and resource limits per shard, that a protocol designer must choose. The choice of these parameters can dramatically impact real-world performance and security of a sharded or parallelized blockchain. This paper will introduce agent-based simulation, which is used in a variety of other fields, as a way for protocol designers and practitioners to overcome some of these problems.

Agent-based simulation is used by practitioners and researchers in algorithmic trading \cite{leal2016rock}, artificial intelligence \cite{silver2017mastering}, autonomous vehicles \cite{bojarski2016end}, cybersecurity \cite{kotenko2010agent}, economics \cite{farmer2009economy}, energy allocation \cite{grozev2005nemsim}, and by the US Commodities and Futures Trading Commission to detect fraudulent market activity \cite{yang2012behavior}. These systems define a set of agents $A_1, \ldots, A_n$ that each have a state space $\mathcal{S}_i$ and interact with each other via a prescribed set of actions $\mathcal{A}$. Each agent receives updates to their state, caused either by the choice of actions of other agents or due to exogenous signals (e.g. external market prices), and then computes a \emph{policy function} $\pi_i$ that selects an action for an agent to take. After running a number of simulations of agents interacting under different ensembles of initial conditions, exogenous data, and policies, a practitioner usually computes statistical averages and/or estimators of game theoretic quantities \cite{lanctot2017unified} to provide estimates of `macroscopic' quantities that are emergent properties of the agent interaction. The statistical sampling of these ensembles is usually implemented via Monte Carlo (MC) and Markov chain Monte Carlo (MCMC) methods, while the designing of policies, state spaces, and action spaces is performed via analytic modeling and more recently, deep reinforcement learning \cite{silver2017mastering}. 

While these techniques have been successfully deployed in practice within other financial disciplines, the use of agent-based simulations for blockchain development has been scant. As far as the authors can tell, agent-based blockchain simulation has focused on refining selfish mining rewards \cite{gobel2016bitcoin}, reducing the energy usage of Proof-of-Work systems \cite{brousmichc2018blockchain}, and on disproving the claims of various block-free ledgers (e.g. IOTA \cite{bottone2018multi}). However, agent-based simulation can be used as a production tool to help protocol designers optimize parameters, estimate latency and bandwidth usage, and to estimate the true cost of security for Proof-of-Stake systems. Moreover, the pervasive usage of agent-based simulations in production environments within algorithmic trading intimates that the technique is useful for monitoring and estimating risk within live blockchain systems.

We aim to illustrate the versatility of agent-based simulation via an analysis of Chainweb, a parallelized Proof-of-Work protocol that has eluded a closed-form security proof due to explicit correlation between chains that makes it difficult to use standard probabilistic tools. Our simulations, which are designed to handle arbitrary distributed consensus protocols, will show that we can get high-fidelity estimates for the latency-security trade-off in Chainweb and an estimate for how much miners lose when they try to censor particular chains.

Finally, we note that estimating the precise cost of security is a difficult challenge that involves a number of variables, some of which are exogenous to the underlying chain like rental markets \cite{bonneau2016buy}, arbitrage opportunities on centralized and decentralized exchanges, and the difficulty of estimating the market impact of attacks. The simulations in this paper illustrate how \emph{endogenous} design choices, such as difficulty adjustment periods or shard correlations, affect practical protocol performance. Our final section will conclude with a description of how simulations can interact with exogenous data (akin to trading strategies) and how we can use simulation to model the effects of volatile exchange prices and derivatives markets on chain security. These uses are increasingly important as the security of Proof-of-Stake protocols is inseparably tied to exogenous data and judicious parameter selection (e.g. slashing rates in live protocols such as Tezos \cite{goodman2014tezos}). Even alternative Sybil resistance mechanisms such as Proof-of-Replication (used in FileCoin \cite{fisch2018poreps}) and privacy data marketplaces \cite{hynes2018demonstration} end up relying on market making and order matching functionality that is best optimized via agent-based simulation.

\section{Simulation Methodology}
We will discuss our simulation methodology in two parts: one that evinces the features of our generic simulation platform and another that describes how we modeled agents within our system.

\subsection{Simulation Platform}
Our \verb=C++= simulation platform consists of a custom discrete event simulation that emulates the peer-to-peer network of a blockchain system and uses a probabilistic generative model to generate events that corresponds to actions that agents can take within a blockchain system. We use MCMC methods to sample from this model to generate the next event and to propagate this event via the network graph. Each protocol, whether it be Proof-of-Work, Proof-of-Stake, of Proof-of-Replication, has a standardized interface to define its generative model. This interface forces the protocol designer to specify a model for event arrival times, a method for selecting an agent or committee to produce an event, and actions that agents are supposed to take. The protocol designer also has to select a routing algorithm that represents how protocol users are gossiping blocks and off-chain packets with each other. We deliberately separate the routing algorithm from the underlying graph of miners and users, allowing us to test how the system behaves with different gossip protocols. This allows for a protocol designer to test their protocol under deterministic algorithms, such as Kademlia \cite{maymounkov2002kademlia}, or a randomized gossip method \cite{boyd2006randomized}. Moreover, the design of this platform is based on a combination of high-frequency financial back-testing simulations, which use highly optimized, low latency versions of probabilistic networking models, and computational physics simulations, which provide optimized methods for evolving systems of interacting objects.

The simulation also includes a domain-specific language for describing the statistical calculations and policies of individual agents. In order to mimic cryptographic simulation-based proofs \cite{lindell2017simulate} and ideal functionalities, we allow for agents to directly interact with the MCMC model. This allows for honest agents to mutate stochastic components of the simulation related to their local state (e.g. hash power, bonded stake or space) and lets Byzantine agents mutate the state of other agents. Agent computations are not restricted and allowed to be generic; for instance, an agent can use a trained TensorFlow or PyTorch model as a policy. Given that there is a modicum, at best, of data stored in blockchains, all agent policies used in the sequel will be rules-based and closer to high precision, low recall trading strategies versus low precision, high recall reinforcement learning policies.

\subsection{Agent Design Methodology}
We use agent-based models to model rational, Byzantine, and adversarial miner strategies. We assume, without the loss of generality, that our agents satisfy
the Byzantine-Altruistic-Rational assumption \cite{aiyer2005bar}. Agents are represented via a state and action space model, in which agents provide a 
utility function $U : \mathcal{S} \rightarrow \R$ that is used to adapt a policy, $\pi: \mathcal{S} \rightarrow \mathcal{A}$, where the state space is $S$ and
the action space is $\mathcal{A}$. In this paper, we define our state space to be $\mathrm{A}_t \times \Delta^{\Nshards}$, where $\mathrm{A}_t$ is the miner's local arboretum at time $t$ 
and $\Delta^{\Nshards}$ represents the miner's hash power distribution. Note that we explicitly exclude block withholding from the action space, as the only choices that a miner 
can make involve changing their hash power distribution. In particular, this also means that our action space is simply $\Delta^{\Nshards}$, as any action 
corresponds to a choice of hash power distribution. As this is preliminary work, we excluded selfish mining and withholding attacks to simplify the statistical analysis\footnote{Our future work will discuss using adaptive sampling methods to deal with the non-stationarity that occurs when adding block withholding} in \S4. 
We will expand upon these results and include selfish mining and withholding attacks in future work. Agents adjust their policies by attempting to optimize their expected reward under a prescribed utility function. 
For simplicity, we assume that the utility function is $C^1$, so that we can optimize it via gradient descent. Drawing inspiration from the reinforcement learning literature, 
we define the expected reward as the $k$-step exponential moving average ($\alpha < 1$) of the utility function:
\[
\Expect_{\alpha}[R_t] = \sum_{\tau = 0}^{k-1} \alpha^{\tau} U(S_{t-\tau})
\]
where $S_t \in \mathcal{S}$ is the state observed by the agent at time $t$. In this paper, $S_t$ is simply the agent's local copy of the Chainweb arboretum
at time $t$ and their current hash power distribution. From here on, we fix $\alpha = \frac{1}{2}$ and let $\Expect[R_t] = \Expect\left[R_t, \frac{1}{2}\right]$. Thus, given a utility function, an agent will update their state via the ordinary gradient descent iteration,
\[
S_{t+1} = S_{t} - \eta \sum_{\tau=0}^{k-1} 2^{-\tau} \nabla_{S_{t-\tau}} U(S_{t-\tau}).
\]
Using gradient descent also helps us enforce a locality constraint that ensures that miner strategies are purely local and not reliant on long-term historical results. 

In order to program the agents within our simulation environment, we have developed a custom domain-specific language that aims to optimize the computation of utilities and policies for each simulated agent. This will be discussed in more detail in a subsequent paper by a subset of the authors. Using this language, we can specify what statistical signals trigger agents to take actions, allowing for us to describe complex agent strategies and policies via a simple script. Our language also allows for us to treat agents as a template, which lets us take a set of nodes and assign the same strategy to that set of nodes. This allows for easy specification of agent distributions, allowing for a user to specify policies $\pi_1, \ldots, \pi_n$ and their relative proportions $p \in \Delta^n$.   

\subsubsection{Utility Function}
The authors of \cite{dhamal2018stochastic} show that in the continuous time and zero latency (e.g. no peer-to-peer network) setting, Markovian agents maximize
their expected reward in Bitcoin by optimizing a utility function propotional to $\frac{R(t)}{H(t)}$, where $R(t)$ is the number of rewards that the agent has 
accrued at time $t$ and $H(t)$ the agent's hash power expenditure at time $t$. Since this paper assumes that $H(t)$ is constant and uniform for all agents, we 
can simply define the single chain utility function to be:
\[
U_{i,\alpha}(S_t) = U_{i,\alpha}(A_t, h_t) = \frac{\gamma(\text{height}(\mathcal{C}_{\alpha, t}))}{h_t}
\]
where $h_t \in (0,1)$ is the fraction of hash power placed on chain $\mathcal{C}_{\alpha,t}$ is the $\alpha$th chain of the arboretum $A_t$ and $\gamma$ is the fraction of blocks in chain $\mathcal{C}$ that agent $i$ produced (see equation \ref{eq:gamma}). Since both values fall in $(0,1)$, we've removed the issue of ensuring that the dynamic ranges of these quantities are compatible and will not under/overflow. In future work, we plan on allowing $h_t$ to be an arbitrary positive real number, representing energy costs, which will force us to add in a scaling constant to this utility function.

In the multiple chain setting, we simply define the $i$th miner's utility function as:
\[
U_i(S_t) = \sum_{\alpha \in [\Nshards]} U_{i,\alpha}(S_t)
\]

\section{Chainweb Details}
For demonstration of the proposed modeling techniques we will analyze Chainweb \cite{kadena2018}, a parallelized Proof-of-Work network architecture designed to provide high security and high transaction throughput. Chainweb serves as an ideal use case for simulation because designers have to choose $\Omega(\Nshards^2)$ parameters and because it is difficult to formally prove that Chainweb is resistant to miner censorship. As there has not been a formal description of Chainweb in the literature, we will provide both an informal description (with visualization) and a formal description that proves that the parallelized system achieves liveness.

\subsection{Informal Description}
\begin{figure}
    \includegraphics[scale=0.30]{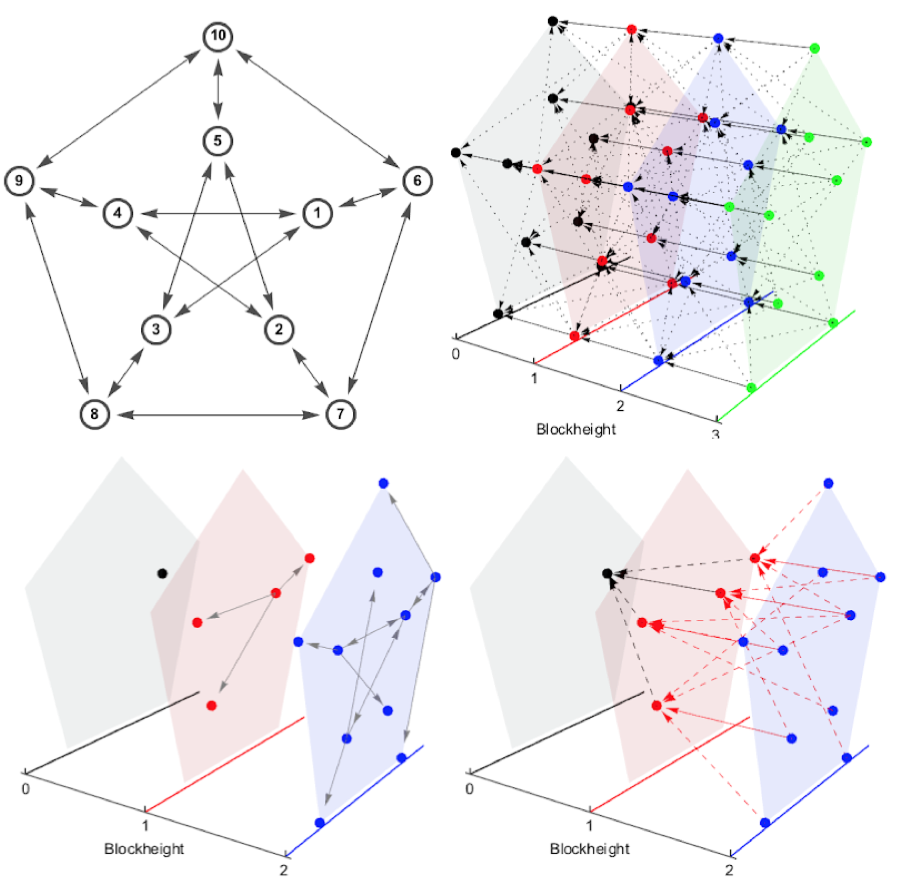}
    \caption{Visualization of Chainweb Base Graph using the Petersen graph as base protocol configuration}
    \label{fig:petersen}
\end{figure}

Chainweb is a parallel chain Proof of Work architecture that combines hundreds or thousands of individually mined peer chains into a single network. Figure \ref{fig:petersen} depicts ten chains that are connected via the Petersen graph. The three-dimensional figures show the dependencies of blocks at different heights on blocks of lower heights. Each chain in the Chainweb braid maintains its own ledger, and blocks on each chain are mined with Nakamoto-style Proof of Work hashing. Peer chains incorporate each other’s Merkle roots to enforce a single super chain that offers an effective hash power that is the sum of the hash rate of each individual chain. Each chain in the network mines the same cryptocurrency which can be transferred cross-chain via a Proof-of-Burn verified with trustless Simple Payment Verification (SPV) at the smart contract level. The configuration of the Chainweb braid and the relationship between chains is fixed at launch and can be hard forked to larger configurations as throughput demands. 
The majority of miners are expected to mine the entire Chainweb braid, and as such each miner will maintain its own local version of Chainweb, the braid of which will be recombined with those of other miners as blocks are created. 

\subsection{Formal Description}
\subsubsection*{Notation}
\begin{itemize}
    \item For any $n\in\N, [n] = \{1,2, \ldots, n\} \subset\N$
    \item $\mathcal{T}_{h, d}$ is the set of a directed trees with maximum height $h$ and
        maximum degree $d$
    \item $\mathcal{T} = \displaystyle{\lim_{h, d \uparrow \infty}} \mathcal{T}_{h, d}$.
        We can take this limit as there is a natural lattice: $\mathcal{T}_{h,d} \subset
        \mathcal{T}_{h', d'}$ if $h < h'$ and $d < d'$
    \item $\mathcal{B} \subset \{0,1\}^*$: The space of admissible blocks
    \item $\GenesisTree$: The tree with one node that represents the genesis block (e.g.
        the empty blockchain). We define $\height(\GenesisTree) = 0$
    \item If we have a graph $G = (V, E)$ define the boundary operator $\partial: V \rightarrow
        2^V$ by $\partial(v) = \{ w: (v,w) \in E\}$
\end{itemize}

\begin{figure}
    \centering
    \includegraphics[scale=0.10]{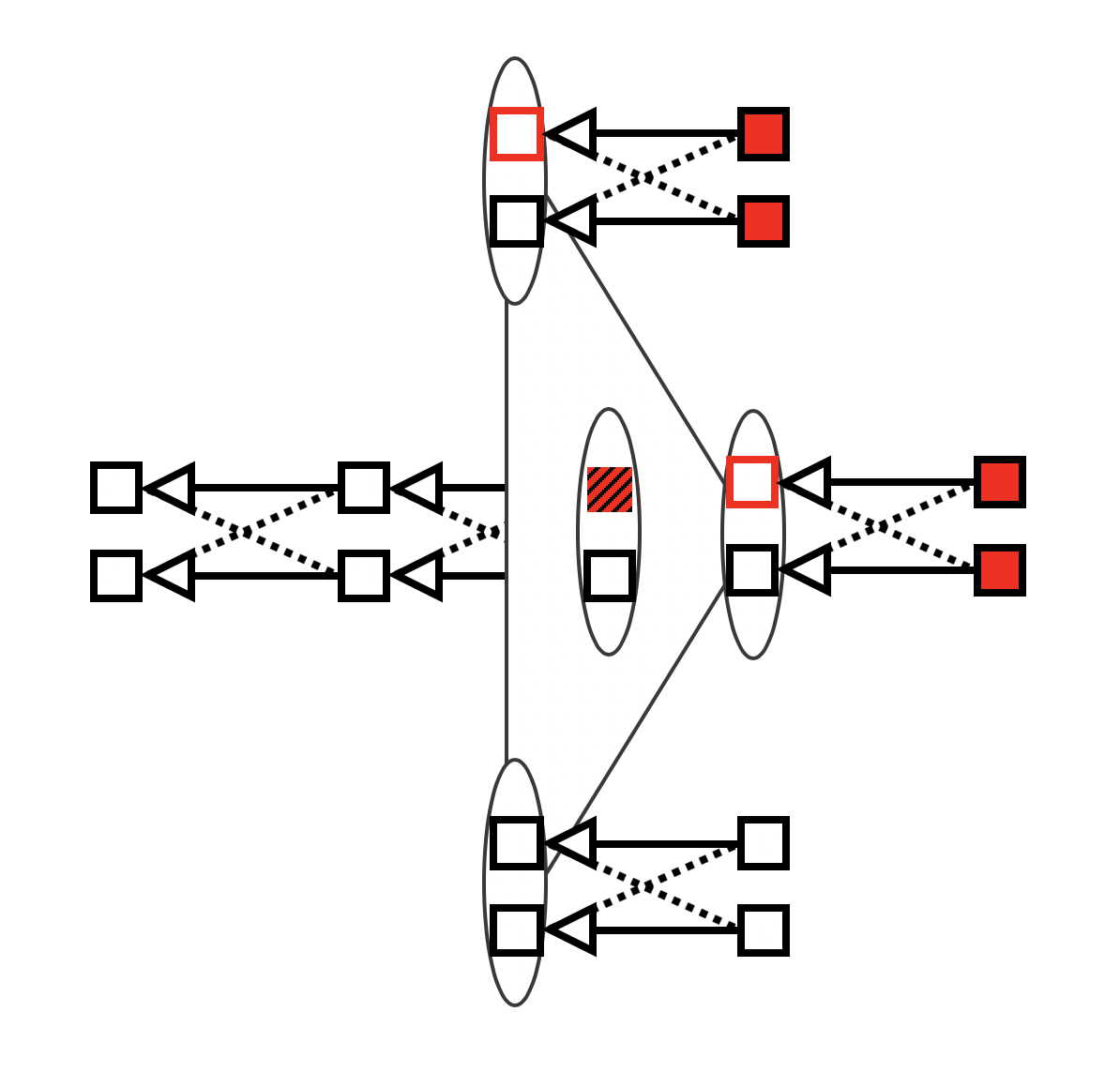}
    \caption{A visualization of the multiple local copies of Chainweb that are maintained by miners, here depicted as ovals, with blocks being generated from left to right and Merkle root references as arrows; this replication generates the arboretum structure. The red block represents two conflicting blocks in the same location, creating a fork which must be reconciled.}
    \label{fig:arboretum}
\end{figure}
\subsection*{From single chain to multiple chains with constraints}
In order to fully-specify a single chain PoW system, one needs to define the following \cite{pass2017analysis}:
\begin{enumerate}
    \item Network graph that describes how miners and users communicate
    \item Merkle trees for each miner that represent each miner's copy of the blockchain. 
    \item A process for dynamically growing and updating the Merkle tree
\end{enumerate}
When we refer to trees, we will refer to a block tree, where each vertex of the tree represents a block. In the multiple chain world, in order to balance security and throughput, we need to
replace the single Merkle tree rooted at each miner with a set of Merkle trees and a set
of constraints that do not let the different chains get too far out of sync with out
other. To formalize this, we will first need to define the terms network graph, base graph,
and arboretum:
\begin{defn}
A \textbf{network graph} is a weighted, undirected graph $\Gminer = (\Vminer, \Eminer,
\Wminer)$, where each vertex $v \in \Vminer$ represents a miner, each edge $e \in \Eminer,
e = (v,w)$ represents a peer-to-peer network connection between miners $v$ and $w$, and
$\Wminer : \Eminer \rightarrow \R$ maps an edge to a latency. An $\Nshards$-\textbf{base graph} is an
unweighted, undirected graph $\Gbase = ([\Nshards], \Ebase)$, where each vertex represents
a separate chain and each edge represents a constraint. 
\end{defn}
In practice, the network graph is changing as users join and leave the network and the edges
change in accordance with the peer-to-peer networking protocol.\footnote{Formally, if we
    think of all possible node sets $N$ as being a subset of $[\Nmax]$, then we can view
    the networking protocol as a map $\rho : 2^{\Nmax} \rightarrow 2^{\binom{\Nmax}{2}}$,
    which maps a node set into an edge set.} In this work, we will consider the network graph
static and unchanging during the remainder of our analysis. We next need to define how our
constraint set, represented by $\Gbase$, affects the allowable Merkle trees.  
\begin{defn}
An \emph{arboretum} $A$ is a triple $(V, E, T)$ where $V \subset \N$ is a vertex set, $E
\subset 2^{\binom{V}{2}}$ is an edge set and $T : V \rightarrow \mathcal{T}$ is an
operator that maps a vertex to a tree. If $\Lambda: E \times \sqcup_{v \in V} T(v) \rightarrow
\{0,1\}$ is a predicate, a $\Lambda$\textbf{-arboretum} is an arboretum that $\Lambda((v,w), T(v)
\sqcup T(v)) = 1$ for all $(v,w) \in E$. Finally, if $G' = (W, F)$ is a subgraph of $G =
(V, E)$, we define a \textbf{subarboretum} centered on $(W,F)$ to be $(W,F,T\vert_{W})$
\end{defn}
An arboretum is distinct from a forest, as a forest does not prescribe
connectivity between trees, whereas the edge set in an arboretum provides a natural graph
distance between any two trees. Figure \ref{fig:arboretum} depicts an arboretum where $\Gminer$ is the triangle graph and $\Gbase$ is a length-1 path with two vertices and one edge. The use of an arboretum is a key feature that is important to Chainweb's performance claims, as the choice of graph can dramatically affect performance in a sharded PoW setting. More precisely, if an edge $e = (v,w) \in \Ebase$, this means that $i$th
block of chain $v$ can only be added if the $i-1$st block of $w$ has been added. Thus, a
valid Chainweb arboretum will need to satisfy the following predicate for all $(v, w) \in E$:
\begin{align}
    &\Chainpred((v, w), T(v) \cup T(w)) = \nonumber \\
                                       &(\height(T(w)) == \height(T(v)) - 1) \nonumber \\ 
                                       &\vee (\height(T(w)) - 1 == \height(T(v))) \nonumber \\
                                       &\vee (\height(T(w)) ==  \height(T(v))) \nonumber \\ 
                                       &\vee (\height(T(v) == 0) \nonumber \\
                                       &\vee (\height(T(v) == 0) \vee \mathsf{compatible}(T(v), T(w))
\end{align}
The last predicate, $\compat(T(v), T(w))$ ensures that the block at the tip of $T(v)$ at height $h$ includes a header from a block at height $h-1$ of $T(w)$. This condition is illustrated in figure \ref{fig:arboretum}, where we see the two blocks in red depending on both their predecessor on the same chain and the adjacent chain. By assumption, we have $\compat(T_{\emptyset}, T_{\emptyset}) = 1$ Finally, given a predicate $\Lambda$, we define a dynamic process on an arboretum that represents a single miner's mining activity:
\begin{algorithm}
    \KwData{An initial arboretum $A_0 \leftarrow ([\Nshards], E, T_0)$, where $T_0(i) = \GenesisTree$, $t \leftarrow 0$}
    \While{true} {
        $\tilde{A}_{t+1} = (\tilde{V}, \tilde{F}, \tilde{T}_{t+1}) \leftarrow$ largest
        $\Lambda$-subarboretum of $A_t$ \;
        $(i, B) \in \tilde{V} \times \mathcal{B} \leftarrow$ Apply mining strategy and find chain $i$ that is
        admissible and mine a valid block $B$ on it (possibly causing a fork)\;
        $T_{t+1} \leftarrow$ Apply block $B$ to $\tilde{T}_{t+1}(i)$
    }
\end{algorithm}

In order to prove that the system achieves liveness, assuming that there is a non-zero amount of hash power on every shard, we need to show that this process 
can continue in an infinite loop without halting for Chainweb. This can happen, for instance, if 
there are no $\Chainpred$-subarboreta of $A_t$. We will show that this cannot happen:
\begin{claim}\label{claim}
    If $G = (V, E)$ is connected, then for all $t \in \N$, there exists a non-empty
    $\Chainpred$-subarboretum of $A_t$
\end{claim}

This claim is proved in the appendix. Let $B_t = (C_t, D_t, \tilde{T}_t)$ be the non-empty $\Chainpred$-subarboretum of $A_t$.
We call $C_t$ the \emph{cut set} of the chain at time $t$. The above claim says that the
cut set is never empty and we can always make progress and the liveness of the set of chains reduces to the liveness of the individual chains. Chainweb thus consists of a miner
graph $G$ and a series of arboreta, one for each vertex, denoted $A_{t,v}$. We note for
completeness that one can also frame Chainweb as an arboretum on the strong graph product
of a network graph and base graph.

\subsection{Implementation of Chainweb}
We represent Chainweb via a graph of miners each of whom holds a copy of their local arboretum. We sample new blocks from a probabilistic model representing Chainweb that has the following parameters:
\begin{itemize}
    \item $\lambda_{\alpha}$: Block frequency for chain $\alpha$
    \item $\mathbf{\zeta}_i \in \Delta^{\Nshards}$: Hash power distribution of miner $i$
    \item $H_i \in \R_{\geq 0}$: Hash power of miner $i$
    \item $H_{\text{total}}$: The total hash power of the system, defined as 
    \[
    H_{\text{total}} = \sum_{i=1}^{\Nminers} H_i
    \]
    \item $\Gamma_{\alpha}$: The fraction of hash power on chain $\alpha$, defined as:
    \[
    \Gamma_{\alpha} = \frac{1}{H_{\text{total}}} \sum_{i=1}^{\Nminers} H_i \zeta_{i,\alpha}
    \]
\end{itemize}
We sample a new block on chain $\alpha$ at time $t$ created by miner $j$, $B_{\alpha,j,t} \in \mathcal{B}$, via the following generative model:
\begin{align*}
    \alpha &\sim \mathsf{Multinomial}(\Gamma_1, \ldots, \Gamma_{\alpha}) \\
    j &\sim \mathsf{Multinomial}(\zeta_{1, \alpha}, \ldots, \zeta_{\Nminers, \alpha}) \\
    t &\sim \mathsf{Poisson}(\lambda_{\alpha})
\end{align*}
Given this model and initial conditions (such as a choice of genesis block and random number seeds), our simulation platform can sample a trajectory of how Chainweb might evolve. Each miner's policy is represented via a function that adapts and optimizes their own hash power distribution, $\mathbf{\zeta}_i$, which allows for the miner to affect the sampled trajectory. We also note that this model shares a number of similarities with multivariate Hawkes processes \cite{embrechts2011multivariate}. In all simulations performed in this paper, we make the following assumptions about the above parameters (as these likely reflect the choices in Chainweb upon launch \cite{kadena2018}):
\begin{itemize}
    \item All chains have the same block production rate: $\exists \lambda \in \mathbb{R}_{\geq 0}\; \forall \alpha \in [\Nshards]\; \lambda_{\alpha} = \lambda$
    \item We use uniform hash power, e.g. $\exists H \in \mathbb{R}_{\geq 0}\; \forall i \in [\Nminers]\; H_i = H$
    \item Our network graph will only contain miners (e.g. we are assuming that all transaction generating participants are also miners)
    \item All simulations in this paper route blocks using either $k$-nearest neighbor routing (like Bitcoin) or use randomized gossip \cite{boyd2006randomized}
    \item All latencies were sampled from the Bitcoin latency distribution in \cite{gencer2018decentralization}
\end{itemize}
Finally, we note that we do not include an explicit difficult adjustment in our simulations, even though it is supported within the platform. While changing the difficulty adjustment time window can have dramatic effects on the profitability of selfish mining \cite{grunspan2018profitability}, we wanted to reduce the noise in our statistics and plan on describing selfish mining optimization in a subsequent work.

\section{Results}
We performed two experiments to validate our simulation methodology and to estimate endogenous costs of security in Chainweb. Our goal is to illustrate how modeling adversarial and networking behavior can help choose design parameters such as $\lambda_i$ and the choice of base graph that is used in the production client.

\subsection{Network Analysis}
Our first experiment aims to test how different network and base graph configurations lead to changes in system dynamics under different routing profiles. As Claim 1 only guarantees that we will eventually achieve liveness in Chainweb provided that all chains have non-trivial chain growth, our goal is to perform a numerical study of how realistic chain growth looks and to find an estimate for how long it takes for a block to reach the whole network. Statistically estimating chain growth and the time to achieve liveness is crucial for deciding how to set parameters the block production rates, $\lambda_i$, as there is a natural trade-off between liveness time and block production rate \cite{sompolinsky2015secure}. Since Chainweb's throughput is bounded by the diameter of the base graph \cite{kadena2018}, we aimed to use graphs that are the best known solutions to the degree-diameter problem. For these experiments, we used $\Nminers = 8192$ and $\Nshards = 57$, with the choice of shards due to the existence of a Moore Graph, the Hoffman-Singleton graph, that solves the degree-diameter problem \cite{miller2005moore}. 

We will describe the statistics of interest. For a miner $i$ and chain $\alpha$, let $h_{i,\alpha}(t) \in \N$ be the height of the longest branch of chain $\alpha$ that miner $i$ has seen. For notational convenience, we assume that $h_{i,\alpha}(t) = 0$ for $t < 0$. Define the \emph{height function}, $H : \R_{\geq 0} \rightarrow \R$ as:
\begin{equation}
H(\tau) = \frac{1}{\Nminers} \frac{1}{\Nshards} \sum_{i=1}^{\Nminers} \sum_{\alpha=1}^{\Nshards} \Expect_{t}[ h_{i,\alpha}(t)) - h_{i,\alpha}(t - \tau)] 
\end{equation}
where $\Expect_t$ is the expectation over all environments up to time\footnote{Our Chainweb model is a composed of Markov models and inherits a natural filtration that $\Expect_t$ is defined on (regardless of initial environment).} $t$. In practice, if we simulate $k$ trajectories until time $T$, we can approximate this expectation as:
\[
\Expect_t[h_{i,\alpha}(t) - h_{i, \alpha}(t-\tau)] \approx \frac{1}{kT} \sum_{t \leq T} (h_{i, \alpha}(t) - h_{i,\alpha}(t - \tau))
\]
Intuitively, this tells us what the expected height change of the blockchain is within a window of size $\tau$. By averaging over windows that start at different times $t$, we can de-noise the variations of this measurement and observe behaviors that are persistent across different initial conditions. Finally, let $\tau_{i,\alpha, k}$ be the time that the $k$th miner receives block $i$ on chain $\alpha$. Let $\hat{\tau}_{i,\alpha,k}$ be the random variable that is $\tau_{i,\alpha,k}$ conditional on block $k$ being on the main branch of chain $\alpha$. Define the \emph{liveness time} $\overline{\tau}$ to be:
\begin{equation}\label{eq:liveness}
\overline{\tau} = \frac{1}{\Nshards} \sum_{i=1}^{\Nshards} \Expect_{k} \left[ \max_{k\in[\Nminers]} \hat{\tau}_{i,j,k} - \min_{k\in [\Nminers]} \hat{\tau}_{i,\alpha, k} \right]
\end{equation}
This measures the average time that it takes for a block that reaches the main chain to reach all miners. If we use randomized gossip, one expects this time to simply be the covering time of random walk on the miner graph, which is controlled by the spectral gap of the miner graph \cite{lovasz1993random}. However, since we are conditioning on blocks that make up the main chain, this number can be significantly greater than the covering time.

\subsubsection{Verification Runs}
\begin{figure}
    \centering
    \includegraphics[scale=0.17]{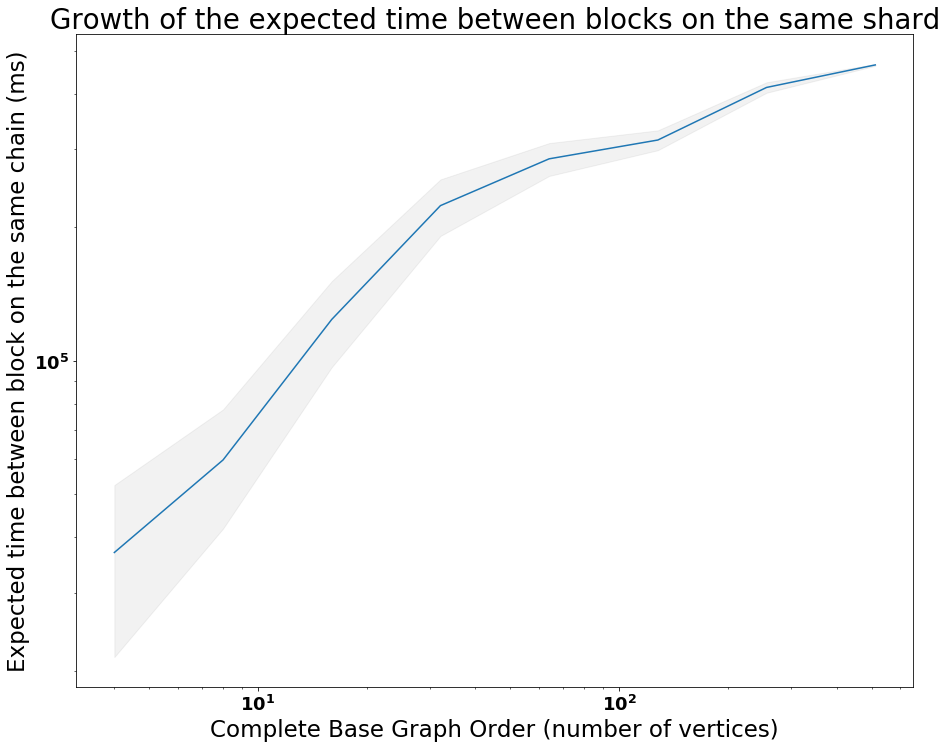}
    \caption{This figure shows the average inter-arrival time (e.g. time between successive blocks on the same chain) as a function of $n = |\Vbase|$ with $\Gbase = K_n$, the complete graph on $n$ vertices. We expect super-linear growth in this time, as each chain has to wait on all other chains before it can make progress. This continues to be true even up to $|\Vbase| = |\Vminer|/2$.}
    \label{fig:verification}
\end{figure}

We performed a few verification runs to show that our simulation is replicating the behavior of Chainweb. For these runs, we let set the base graph be equal to the complete graph, $\Gbase = K_{\Nshards}$. In this situation, since every shard depends on every other shard, we expect the time between successive blocks on the same chain to have super-linear scaling in the size of the base graph. We ran 100 simulations with 16,384 miners and a base graph with $|\Vbase| \in \{2^k : 1 \leq k \leq 13 \}$ and calculated this time. In figure \ref{fig:verification}, we see super-linear scaling (note that both scales are logarithmic) and decreasing variance as we increase the base graph size. We also note that one expected some finite-size effects as $|\Vbase| \approx |\Vminer|$, which is the likely cause of the decay in the rate of growth of this curve.

\subsubsection{Liveness Time}
\begin{figure}
    \centering
    \includegraphics[scale=0.17]{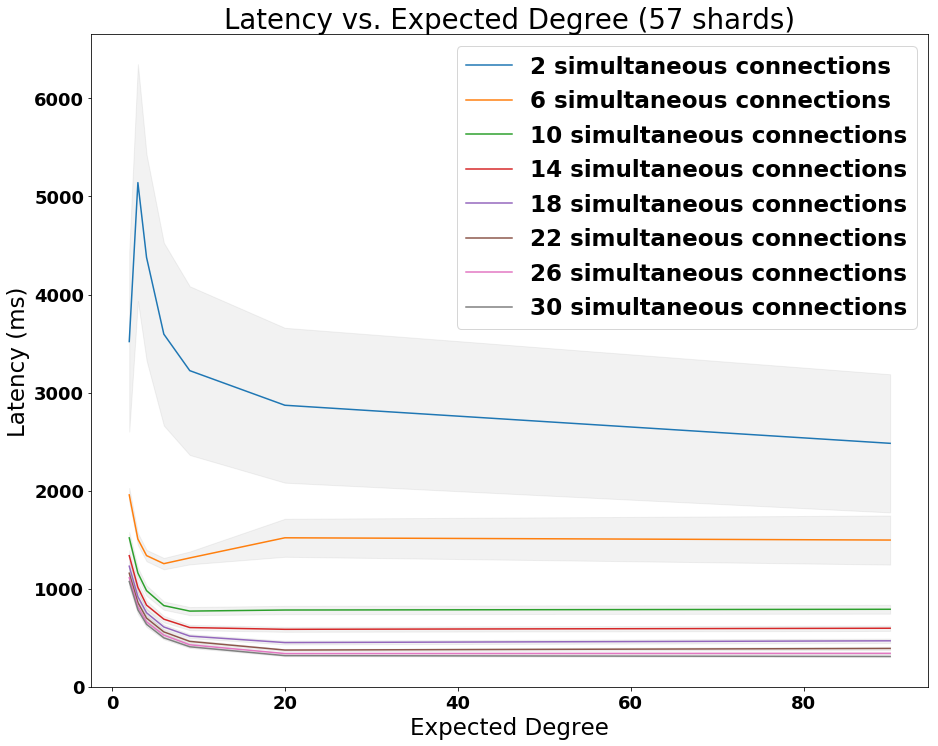}
    \caption{In this figure, which plots expected latency (and confidence intervals, represented via error bands) against expected degree of a Barab\'si graph, we see that by increasing the bandwidth used, we quickly saturate the graph.}
    \label{fig:liveness_time}
\end{figure}

To assess liveness time (\ref{eq:liveness}) under realistic conditions, we constructed a realistic internet graph, the Barab\'asi small world graph \cite{albert1999internet}, used a randomized gossip protocol \cite{boyd2006randomized}, and sampled latencies the Bitcoin latency distribution \cite{gencer2018decentralization}. In order to stress the system, we assumed a block production rate of $\lambda = 1 \text{Hz}$. In figure \ref{fig:liveness_time}, we see a plot of the expected degree of a Barab\'asi graph versus liveness time. We used a miner graph of size 32,768 and used the Hoffman-Singleton graph as a base graph to generate these figures. The different curves correspond to a different number of simultaneous connections, which is how we quantify bandwidth. This corresponds to the upper bound on the number of neighbors forwarded to by a miner. We can see that by the time we get to 10 simultaneous connections, we are close to saturating the high-bandwidth limit, even for low-connectivity Barab\'asi graphs. Data of this form helps protocol developers assess design decisions and choose peer-to-peer networking capabilities whose bandwidth-latency tradeoff matches the expectations of the consensus mechanism.

\subsubsection{Random Miner Graph}
In order to assess how Chainweb performs under more extreme network graphs, we looked at a modified version of Erd\"os-Renyi random graphs $G_p$ such that if $i \in [\Nminers-1], j \in [\Nminers] - \{i +1\}$, then the edge $(i,j)$ is included with probability $p$ and such that the edge $(i,i+1)$ is always included. We chose this ensemble because it ensures that our graph is connected (via the inclusion of the line edges $(i, i+1)$), but still inherits the Erd\"os-Renyi phase transition. In these experiments, we set the block production rate to be that of Bitcoin ($\lambda = \frac{1}{600}\text{Hz}$) so that these results could be interpreted in terms of real world data. Moreover, we averaged over 10 instances of each random graph $G_p$ and used the Hoffman-Singleton graph as a base graph. 

\begin{figure}
    \includegraphics[scale=0.45]{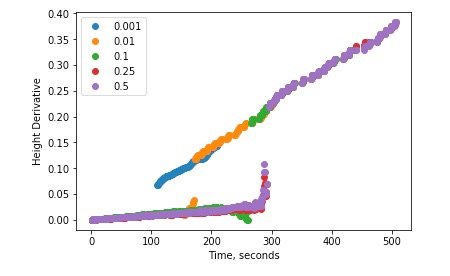}
    \caption{Computation of $H(\tau)$ (denoted height derivative) for times within one block interval ($\lambda = \frac{1}{600}\text{Hz}$.) The legend indicates the Erd\"os-Renyi probability $p$ that was used to generate $G_p$.}
    \label{fig:height_deriv}
\end{figure}

In Figure \ref{fig:height_deriv}, we see computations of $H(\tau)$ within one block interval. We see that for high values of $p$ (e.g. the miner graph is more connected), there is a sharp uptick in expected height increase around a half block time (300s). As we decrease connectivity, we expect there to be a more gradual height increase as some of the graph will have received the latest block, whereas other are still waiting to receive it and/or are on other forks. This is the expected behavior: when height increases propagate uniformly and rapidly throughout the network graph, we expect a sharp uptick in expected height change, as most participants receive the block (on average) at the same time (since miners have the same hash power). However, it is curious to note that the Erd\"os-Renyi phase transition \cite{van2009random} takes place around $p=0.001$ for our graphs and yet we only observe the signs of the transition (e.g. the sharp derivative for $p \in \{0.1, 0.25, 0.5\}$) far away from it. This suggests that the presence of forks can dramatically slow down block propagation and adds in a non-linear latency effect. This effect can likely be measured by miners, who measure their deviation from these expected curves, who can potentially use this information to boost the rewards from selfish mining, akin to latency arbitrage in high-frequency trading. 

\subsection{Adversarial Censorship}
\begin{figure}
    \includegraphics[scale=0.45]{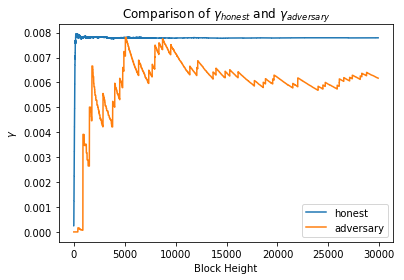}
    \caption{In this image, we see that the expected fraction of honest blocks, $\gamma_{\text{honest}}(h)$ stabilizes rapidly, whereas $\gamma_{\text{adversary}}(h)$ fluctuates wildly, with excursions to the stable point that is achieved by honest agents. This data is taken from a single simulation.}
    \label{fig:single_trajectory_rewards}
\end{figure}

In order to test how resilient Chainweb is to miners attempting to censor a chain $\alpha$ by not allocating any hash power to $\alpha$, we used our domain specific language to construct two agents that do the following:
\begin{enumerate}
    \item Honest Agent: Computes utility and gradient on all chains and updates its hash power distribution via gradient descent
    \item Adversarial Agent: Computes utility and gradient on all chains, updates its hash power via gradient descent, sets the hash power allocated to the censored chain to zero, and then redistributes the excess hash power equally.   
\end{enumerate}
We chose a sequence of twenty five evenly-spaced adversary fractions $f_i \in (0, 1)$ and for each of these, ran 256 simulations (with the Hoffman-Singleton base graph) of 30,000 blocks produced with a $(1-f_i)$ fraction of honest agents and $f_i$ censoring agents. For each miner $i$ and chain $\alpha$, we computed the fraction $\gamma_{i,\alpha}(h)$ of blocks on the main branch of chain $\alpha$ that were created by miner $i$ at block height $h$, where $\forall \alpha \in [\Nshards]\,\forall h \in \N\, \sum_{i \in [\Nminers]} \gamma_{i,\alpha}(h) = 1$. For simplicity, we assume that the block reward is constant and is equal to one coin per block throughout our simulations (as all of our plots are about relative rewards, this choice of units does not affect our results). We define the following two quantities:
\begin{align}
    \gamma_{\text{honest}}(h) &= \frac{1}{|S_{\text{honest}}|} \frac{1}{\Nshards} \sum_{i \in S_{\text{honest}}} \gamma_{i,\alpha}(h) \label{eq:gamma} \\
    \gamma_{\text{adversary}}(h) &= \frac{1}{|S_{\text{adversary}}|} \frac{1}{\Nshards} \sum_{i \in S_{\text{adversary}}} \gamma_{i,\alpha}(h)
\end{align}
where $S_{\text{honest}} \subset [\Nminers]$ is the set of honest miners and $S_{\text{adversary}} = [\Nminers] - S_{\text{honest}}$. In Figure \ref{fig:single_trajectory_rewards}, we see the plot for a single trajectory with $f_i = 0.1$, where we see that $\gamma_{\text{honest}} > \gamma_{\text{adversary}}$ and that the honest curve is much smoother than the noisy adversary curve.

\begin{figure}
    \centering
    \includegraphics[scale=0.17]{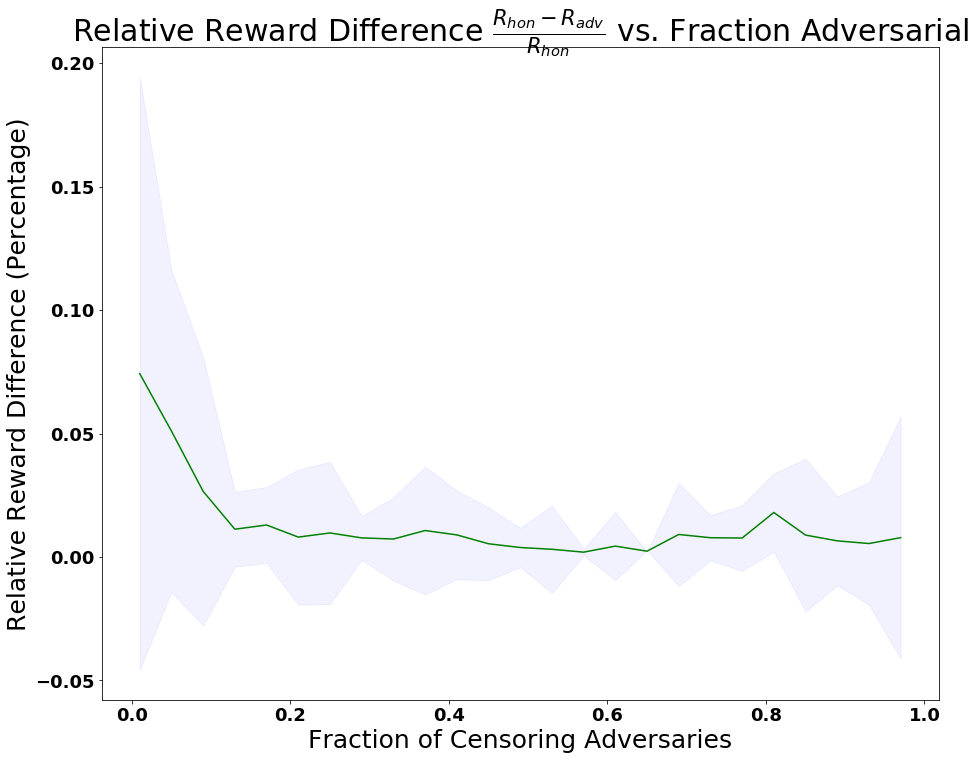}
    \includegraphics[scale=0.17]{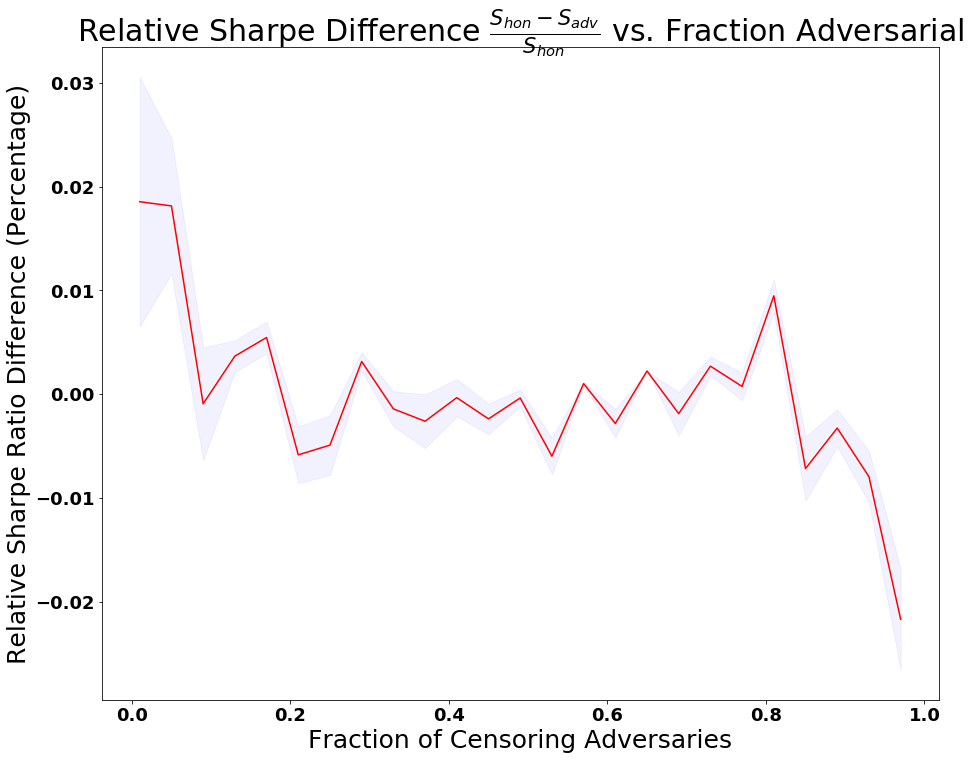}
    \caption{In the top figure, we see the relative expected reward, $\frac{\Expect[R_{hon}]-\Expect[R_{adv}]}{\Expect[R_{adv}]}$ between expected adversary rewards and honest rewards. We see that the honest miner advantage decays quickly from 7\% to 1\%, although honesty appears to be the dominant strategy. On the other hand, the risk adjusted returns, measured via the Sharpe Ratio $S_{\alpha} = \frac{\Expect[R_{\alpha}]}{\sqrt{\Var[R_{\alpha}}}$, $\alpha \in \{hon, adv\}$, has a stronger transition around 80.}
    \label{fig:deltas}
\end{figure}

In Figure \ref{fig:deltas}, we see how the expected difference in rewards and risk-adjusted rewards vary as a function of $f \in (0,1)$. We measure risk-adjusted rewards using the Sharpe Ratio, a common measure of trading strategy performance, which is measured via the mean reward (over all trajectories sampled) divided by the standard deviation of the reward. As the plots illustrate, honest, profit-maximizing agents beat out the censoring agents until 50-60\%, at which point the censoring adversary is increasingly favored.

From this data, we can conclude that it appears to be hard for miners that are aiming to censor a chain to successfully do it as long as there are enough rational, profit-maximizing miners in the system.

\section{Conclusions and Further Work}
We explored how adversarial, agent-based simulation can be used to assess claims and measure networking behavior for a blockchain protocol. Our techniques helped us assess different design choices, such as how the choice of base graph affects liveness time, and we were able to statistically estimate the rewards sacrificed by an adversarial, censoring agent. These techniques helped us evaluate the safety of Chainweb and help us statistically justify that degree-diameter minimizing base graphs prevent censorship attacks. As our experiments with liveness time show, one can use these techniques to optimize peer-to-peer networking performance based on various metrics, such as the number of simultaneous connection in \ref{fig:liveness_time}. Finally, we note that more complicated consensus protocols, such as Proof-of-Stake and Proof-of-Space tend to have a variety of other parameters (slashing, market making or auction parameters, etc.) and these can be optimized in a similar fashion, given a certain mixture of honest and adversarial agents. Our future work will include an analysis of block withholding and selfish mining and analyzing Chainweb (once it is live) by incorporating \emph{exogenous} data to estimate the true cost of security and selfish mining, conditional on the existence of active fiat and derivatives markets.

\section{Acknowledgments}
The authors would like to thank Joseph Bonneau, Yi Sun, Emily Pillmore, and the anonymous reviewers for insightful and constructive comments.

\bibliographystyle{unsrt}
\bibliography{main}

\section{Appendix: Proof of Claim \ref{claim}}
\begin{proof}
    We will prove this by induction. By construction, $A_0$ is a $\Chainpred$-arboretum
    since $\height(\GenesisTree) = 0$ and $\compat(\GenesisTree, \GenesisTree) = 1$. 
    Now assume the induction hypothesis that $A_{t-1}$ has a $\Chainpred$-subarboretum 
    and suppose for a contradiction that $A_t$ does not have a $\Chainpred$-arboretum. 
    We first note that if $\compat(T_t(v), T_t(w))$ is not satisfied, but all of the height
    conditions are satisfied, then one can start a fork at $T_{t-1}(v)$ or $T_{t-1}(w)$ that
    forces compatibility. By assumption $T_{t-1}(v)$ and $T_{t-1}(w)$ are admissible, 
    so this is possible. Thus, we can assume that the compatibility condition is satisfied. 
    Since $A_{t-1}$ has a $\Chainpred$-subarboretum, $\tilde{A}_t$
    was non-empty. Let $i \in [\Nshards]$ be the index of the block mined in the second
    step of the loop. From the definition of $\Chainpred$, this means that
    $\height(T_{t-1}(i)) \in \{ \height(T_{t-1}(j)) + \eta : j \in \partial(i_t), \eta \in \{0,
    1\}\}$ and this set is non-empty since $G$ is connected. We have two cases:
    \begin{enumerate}
        \item $\height(T_{t-1}(i)) = \height(T_{t-1}(j))$: This implies that
            $\height(T_t(i)) = \height(T_t(j)) + 1$, so $j$ is admissible in the next
            round
        \item $\height(T_{t-1}(i)) = \height(T_{t-1}(j)) + 1$: This implies that
            $\height(T_t(i)) = \height(T_t(j)$ in the next round, so both $i, j$ are
            admissible
    \end{enumerate}
    Thus $j$ is admissible and there must be a non-empty $\Chainpred$-arboretum
\end{proof}

\end{document}